\documentclass[aps,prl,twocolumn,showpacs,superscriptaddress]{revtex4}

\usepackage{graphicx}
\usepackage{epsfig}

\usepackage{color}

\begin{document}

\title{Experimental demonstration of continuous variable purification of squeezed states}

\author{Alexander Franzen}
\affiliation{Institut f\"ur Gravitationsphysik,
Universit\"at Hannover and Max-Planck-Institut f\"ur
Gravitationsphysik (Albert-Einstein-Institut), Callinstr. 38,
30167 Hannover, Germany}

\author{Boris Hage}
\affiliation{Institut f\"ur Gravitationsphysik,
Universit\"at Hannover and Max-Planck-Institut f\"ur
Gravitationsphysik (Albert-Einstein-Institut), Callinstr. 38,
30167 Hannover, Germany}

\author{James DiGuglielmo}
\affiliation{Institut f\"ur Gravitationsphysik,
Universit\"at Hannover and Max-Planck-Institut f\"ur
Gravitationsphysik (Albert-Einstein-Institut), Callinstr. 38,
30167 Hannover, Germany}

\author{Jarom\'{\i}r Fiur\'{a}\v{s}ek}
\affiliation{Department of Optics, Palack\'{y} University, 
17. listopadu 50, 77200 Olomouc, Czech Republic}

\author{Roman Schnabel}
\affiliation{Institut f\"ur Gravitationsphysik,
Universit\"at Hannover and Max-Planck-Institut f\"ur
Gravitationsphysik (Albert-Einstein-Institut), Callinstr. 38,
30167 Hannover, Germany}

\date{\today}


\begin{abstract}
We report on the first experimental demonstration of purification of nonclassical continous variable states. The protocol uses two copies of phase-diffused states overlapped on a beam splitter and provides Gaussified, less mixed states with the degree of squeezing improved. The protocol uses only linear optical devices such as beam splitters and homodyne detection, thereby proving these optical elements can be used for successful purification of this type of state decoherence which occurs in optical transmission channels.
\end{abstract}

\pacs{03.67.-a, 42.50.-p, 03.65.Ud}
\maketitle


Purification of quantum states is one of the key techniques for quantum information processing and for long distance quantum communication networks with the purpose to counteract decoherence and to regain the nonclassical properties of Fock states, entangled single photon states as well as squeezed and entangled continuous variable states \cite{BBPSSW96, DEJMPS96}.
First purification protocols for single photon states have been demonstrated experimentally, relying on coincidence measurements, destructive post-selection, and photon counters 
\cite{singlephoton}.

Continuous variable states are of particular interest. The subclass of so-called Gaussian states
are easily accessible experimentally and still provide the resource for many quantum-information protocols, such as quantum teleportation \cite{{FSBFKP98,BTBSRBSL03}}, entanglement swapping \cite{TYAF05}, and dense coding \cite{JZYZXP03}.
They also provide a resource to improve high precision optical measurements, e.g. in gravitational-wave detection \cite{VCHFDS05}.
Therefore much attention has been attracted by the discovery that entangled Gaussian states can not be purified with local Gaussian operations \cite{purification02}. Whereas the Gaussian setting solely builds on parametric oscillation, beam splitters, phase shifters and homodyne detectors, additional non-Gaussian operations like single-photon   detection seemed to be  essential for the purification.
In a more general investigation it was then theoretically shown that the Gaussian setting has to be left at a single point to indeed break this no-go theorem \cite{{BESP03,EBSP04}}. 

Non-Gaussian states can be purified by an iterative Gaussification protocol whose single step involves a mixing of two copies of the state on  balanced beam splitters and conditioning on  Gaussian measurements on one of the output copies
\cite{BESP03,EBSP04,FMFS06,EPBSF06}. In a recent paper \cite{FMFS06} it was shown that an experimentally feasible purification protocol of this kind can counteract the effect of phase fluctuations that occur when squeezed or entangled states are generated or transmitted through a noisy channel.
The outstanding advantage of the protocol considered in \cite{FMFS06} lies in the fact that 
homodyne detection which can be performed with very high efficiency is used as 
a conditioning measurement.
It has been proven that the protocol can be applied iteratively and meets all criteria of a proper purification protocol. In particular it has been numerically shown that the logarithmic negativity \cite{VWe02} of phase diffused entangled states indeed increases through the purification protocol.

In this Letter we report the experimental demonstration of a purification protocol as proposed in \cite{FMFS06}. Two copies of single-mode squeezed states were sent through two channels that exhibited uncorrelated quasi-random phase noise and subsequently 
they were purified. It was especially shown that the protocol can restore squeezing that was \emph{completely} lost due to the phase noise, as suggested by the theory.
To counteract optical loss which is another decoherence effect has not been subject of the work presented here.


\begin{figure}[h!!!!]
\includegraphics[width=7 cm]{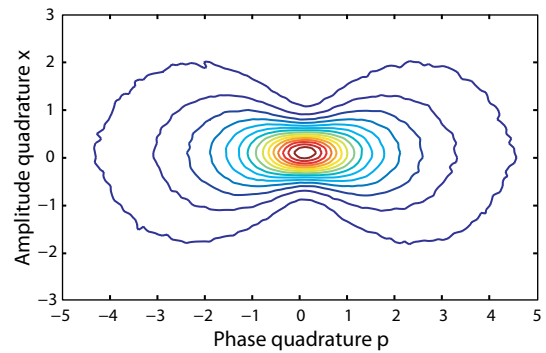}
\caption{(Color online) Numerically simulated Wigner function for a phase diffused squeezed state (in shot-noise units). Here we assumed 10 dB squeezing, an initially pure state and Gaussian phase fluctuations with standard deviation of $\sigma=0.523$ ($\approx\ \textrm{30}^{\circ} $), providing a mixed state with clearly visible non-Gaussian statistics.
}
\label{Fig1:Wigner}
\end{figure}
Consider a single-mode squeezed state whose Wigner function reads 
\begin{equation}
W_{\mathrm{SMS}}=(2\pi\sqrt{V_x V_p})^{-1}\exp\left[-\frac{x^2}{2V_x}-\frac{p^2}{2V_p}\right] ,
\end{equation}
where $V_x$ and $V_p$ 
are variances of the $x$ and $p$ quadratures, respectively, with $V_x V_p \geq 1/4$.
Assuming $V_x \leq V_p$ the state is squeezed if $V_x < 1/2$.
Under the influence of the random phase fluctuations this state will evolve into a
mixed generally non-Gaussian state 
with Wigner function given by
\begin{equation}
W(x,p)=\frac{1}{2\pi\sqrt{V_x V_p}}  \int
\exp\left[-\frac{x_\phi^2}{2V_x}-\frac{p_\phi^2}{2V_p}\right]
\Phi(\phi) d\phi,
\label{Wignersqueezed}
\end{equation}
where $x_\phi=x\cos\phi+p\sin\phi$, $p_\phi=p\cos\phi-x\sin\phi$, and
$\Phi(\phi)$ denotes the probability distribution of the random phase shift, 
$\int \Phi(\phi) d\phi=1$. 
The purification protocol as proposed in \cite{FMFS06} is largely independent of the particular form of the fluctuations
and qualitatively holds for any $\Phi(\phi)$.
However, in this work we realized Gaussian phase fluctuations of the form 
$\Phi(\phi)=(2\pi \sigma^2)^{-\frac{1}{2}} \exp(-\frac{\phi^2}{2\sigma^2})$,
being completely described by the single parameter $\sigma$.
Fig.~\,\ref{Fig1:Wigner} shows the calculated Wigner function of a squeezed state with random Gaussian phase fluctuations. It is clearly visible that phase fluctuations transform pure Gaussian 
squeezed states into mixed states with a reduced degree of squeezing. 
However, one also realizes that the phase diffused state exhibits a non-Gaussian Wigner function.
Phase fluctuations therefore naturally lead to a non-Gaussian setting and the no-go theorem for purification no longer holds. In Fig.~\,\ref{Fig1:Wigner} the strength of the fluctuation ($\sigma=0.523$) and the squeezing assumed (10 dB) was higher than in our experiment to highlight the formation of the non-Gaussian Wigner function.


\begin{figure}[h!!!!]
\includegraphics[width=7 cm]{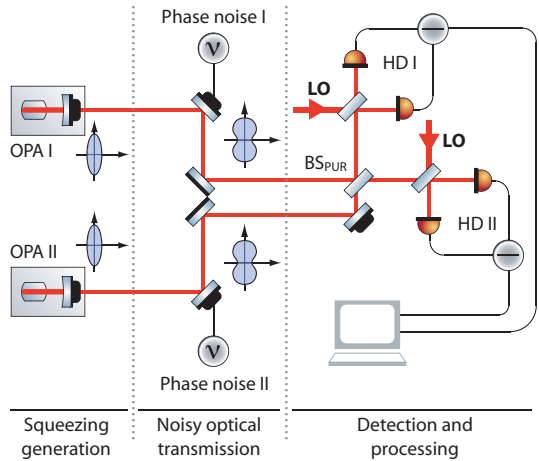}
\caption{(Color online) Schematic experimental setup. Two amplitude squeezed beams were produced by two length-controlled OPA-cavities. Quasi random band-limited Gaussian white phase-noise is applied to both squeezed beams to mimic phase noise in optical transmission channels. The two beams are subsequently overlapped at a 50/50 beamsplitter (BS$_{\mathrm{PUR}}$). 
Time-series of quadrature values for both beam splitter outputs were recorded using balanced homodyne detection (HD) and a data acquisition system.
OPA: optical parametric amplifier; LO: local oscillator; \vrule width 3mm depth -.6mm\,: piezo-electric transducer.
}
\label{Fig2:setup}
\end{figure}

Fig.\,\ref{Fig2:setup} shows the experimental setup that was used to demonstrate purification of phase diffused squeezed states.
The laser source was a continuous wave non-planar Nd:YAG ring laser with 300\,mW of output power at 1064\,nm and 800\,mW at 532\,nm. 
The latter was used to pump two optical parametric amplifiers (OPAs) to produce two amplitude-squeezed light beams with an approximate power of  0.06\,mW at 1064\,nm. Both OPAs were constructed from type I non-critically phase-matched $\textrm{MgO:LiNbO}_3$ crystals inside hemilithic resonators, similar to the design that previously has been used in  \cite{CVHFLDS05}. 
Each resonator was formed by a high-reflection (HR) coated crystal surface (reflectivity $r^2\!>\!0.999$) and a metal-spacer mounted outcoupling mirror (reflectivity $r^2\!=\!0.957$). The intra-cavity crystal-surface was anti-reflection coated for both the fundamental wavelength (1064\,nm, $r^2\!<\!0.05$\,\%) and the second harmonic (532\,nm, $r^2\!<\!0.2$\,\%). The outcoupling mirror had a small reflectivity of $r^2\!=\!0.15\pm0.02$ for 532\,nm. The OPAs were seeded through the HR-surface with a coherent laser beam of 15\,mW power and pumped through the outcoupling mirror with 100\,mW (not shown in Fig.\,\ref{Fig2:setup}), resulting in a classical gain of about 10. 
The lengths of both OPA cavities as well as the phase of the second harmonic pump beams were controlled using radio-frequency modulation/demodulation techniques. All control error-signals were derived from the seed fields reflected from the OPA cavities. 
A nonclassical noise power reduction of slightly more than 3.5\,dB was directly observed with a homodyne detector in combination with a spectrum analyzer. 
Subtracting the variance of the electronic noise yielded squeezing of up to 5\,dB.
The squeezing spectrum was almost white from approximately  5\,MHz to 15\,MHz. 
%
%
Both squeezed fields freely propagated from the OPAs passing high-reflection mirrors that were quasi-randomly moved by piezo-electric transducers (PZT) to mimic the effects of noisy optical transmission channels. The voltages applied to the PZTs were produced as follows. 
Two independent random number generators produced data strings with Gaussian distribution. Both strings were digitally filtered to limit the frequency band to 2--2.5\,kHz.
The output interface was a common PC sound card with SNR of -110\,dB.

Homodyne detection confirmed that for both beams squeezing degraded in the same way when phase noise was increased. 
In order to demonstrate the purification protocol the two squeezed beams were subsequently overlapped at a 50/50 beamsplitter (labeled BS$_{\mathrm{PUR}}$ in Fig.\,\ref{Fig2:setup}). 
The visibility of the squeezed beams' mode matching was 97.4\% and was limited by OPA crystal inhomogeneities.
The phase of the two squeezed beams at the purification beamsplitter was controlled at frequencies below the frequency band of the impressed phase noise. 
The output fields from the purification beamsplitter were then sent to two homodyne-detectors as shown in Fig.\,\ref{Fig2:setup}.
Homodyne-visibilities were at 93.5\,\% for homodyne detector HD\,I and 96.2\,\% for HD\,II. 
Both homodyne detectors were servo-controlled (with loop bandwidth smaller than the noise frequencies) to detect the amplitude quadratures.
%
The detector difference currents were electronically mixed with a 7\,MHz local oscillator. The demodulated signals were then filtered with steep low-pass filter at 30\,kHz and synchronously sampled with 100\,kHz.

Both time series of quadrature values $x_{\mathrm{I}}$ and $x_{\mathrm{II}}$ from the two homodyne detectors were post-processed to accomplish and confirm the purification protocol. The quadrature values $x_{\mathrm{I}}$ measured at the first homodyne-detector (HD\,I) were used to produce a trigger signal for squeezing purification on the second beam. A trigger signal was simply given when a measured quadrature value $|x_{\mathrm{I}}|$ was below some chosen threshold $X$. The second homodyne detector was not part of the purification scheme itself but served as a verification device, confirming that purification had actually occured.

\begin{figure}[h!!!!]
\includegraphics[width=8 cm]{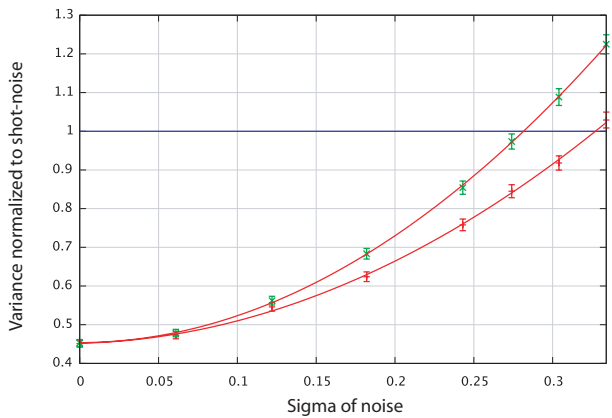}
\caption{(Color online) Amplitude quadrature variance of the single copy squeezed field versus additional phase noise strength (upper curve). Squeezing of 3.55\,dB was directly observed without impressed phase noise ($\sigma\!=\!0$) and completely lost for high levels of phase noise ($\sigma\!>\!0.28$). The lower curve illustrates the amount of squeezing after purification. As predicted by the theory purification becomes more efficient for higher values of $\sigma$. For $\sigma$ values of about $0.3$  squeezing was purified from a field that showed a higher than vacuum noise variance before.
}
\label{Fig3:noise}
\end{figure}

Fig.\,\ref{Fig3:noise} shows the success of our purification protocol versus phase noise strength. Measurement points on the upper curve represent quadrature variance of a single copy before purification, whereas the lower curve provides the measured variances of $x_{\mathrm{II}}$ after two-copy purification. In agreement with the theory, purification becomes more efficient for larger standard deviations of the phase noise $\sigma$. 
The variances plotted correspond to directly observed squeezing; electronic dark noise of the detector was not taken into account. For no additional phase noise ($\sigma\!=\!0$) slightly more than 3.5\,dB of squeezing was observed which was completely lost for noise with $\sigma\!>\!0.28$.

To quantify how effective the purification protocol was the variance of the purified states needs to be compared with the variance of a single copy before purification. If the two states are not copies with identical variances, such a situation might appear in realistic experiments, then the lower variance provides the reference.
In Fig.\,\ref{Fig3:noise} the upper curve represents the variances of both individual copies which were identical within the measurement error bars given, nevertheless
the purified state exhibits lower variance. 


\begin{figure}[h!!!!]
\includegraphics[width=8 cm]{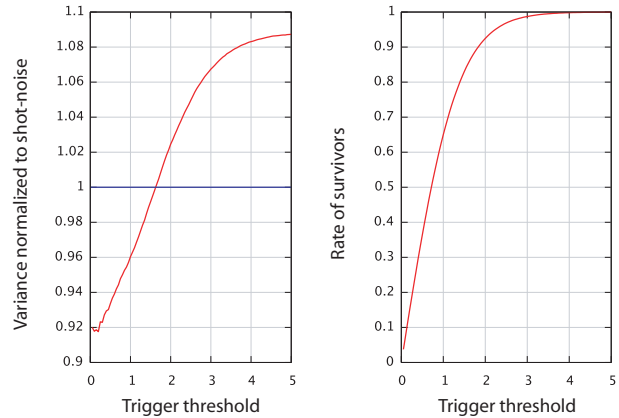}
\caption{(Color online) Purified output variance (left graph) and the rate of surviving data points (right graph) versus trigger threshold $X$. Both curves correspond to $\sigma\!=\!0.304$.
}
\label{Fig4:survivors}
\end{figure}
Fig.\,\ref{Fig4:survivors} further characterizes our purification experiment focusing on the setting with a phase noise of standard deviation $\sigma\!=\!0.304$. It is shown how the purified output variance (left graph) and the rate of surviving data points (right graph) depend on trigger threshold $X$. For a trigger threshold that retains still 50\% of the data, a noise variance above vacuum noise was purified to a value clearly below vacuum noise reference.

\begin{figure}[h!!!!]
\includegraphics[width=8 cm]{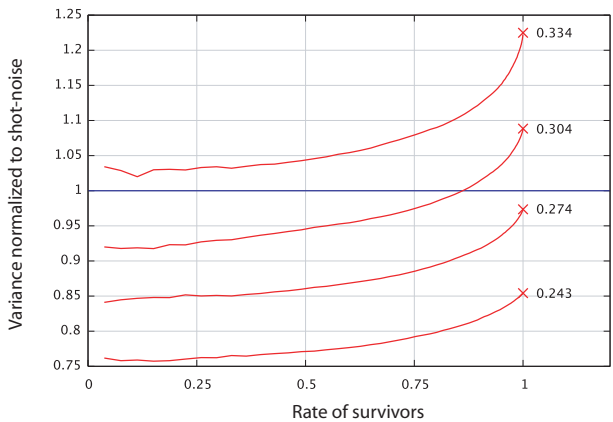}
\caption{(Color online) Measured amplitude quadrature variances of purified states versus state survival rate for  
four different strengths of phase-noise $\sigma$. The shot-noise reference is indicated by the horizontal line. The ripples at small survival rates originate from correspondingly larger uncertainty in the variances obtained.}
\label{Fig5:triggerthreshold}
\end{figure}
Fig.\,\ref{Fig5:triggerthreshold} shows how the variance depends on the rate of retained data points.
Without any purification no state is discarded and 100\,\% of data points are still available.
In all cases choosing the threshold such that half the states were kept provided a significantly reduced variance. 

Successful purification should provide not only an improved strength of nonclassical correlation and Gaussification but also a decreased mixedness of the states. Both effects were clearly observed in our experiment.  
Consider again the purification experiment with phase noise 0.304 in Fig.\,\ref{Fig4:survivors}.
Without purification the product of the amplitude and phase quadrature variances was 7.6, 
after purification we measured 6.0.
Lastly, fitting Gaussian profiles to the measured data manifested a reduction in the reduced $\chi^2$ value by a factor of 4.
Increased squeezing strength, Gaussification and decreased mixedness can physically all be well understood by realizing that the two output fields from the purification beam splitter are correlated and (possibly) entangled. 
Two squeezed fields overlapped on a 50/50 beam splitter have been used before to generate entangled states \cite{OPKP92,Silberhorn01,BSLR03}.
For optimal entanglement generation the phase between the two squeezed inputs is adjusted in such a way that the Wigner functions of the beam splitter outputs are circular. 
In case of the purification of phase diffused squeezed states, the situation is different. The Wigner functions of both beam splitter outputs are non-Gaussian, 
as shown in Fig.~\,\ref{Fig1:Wigner}.
If a high amplitude quadrature value $x_{\mathrm{I}}$ is measured, it is likely that also $x_{\mathrm{II}}$ exhibits a large value.
Exactly those events are discarded in the purification protocol with an appropriately adjusted trigger threshold.
Since such events are 
likely to represent the outer parts of the two convexities, the effect of the random phase fluctuations with all its consequences is counteracted.  
Note that the purified state need not be monitored by HD II and may remain available for further use.



In conclusion we have successfully demonstrated a continuous variable purification protocol using squeezed states which was able to suppress non-Gaussian sources of decoherence such as phase-fluctuations.
To the best of our knowledge this is the first experimental two-copy purification of nonclassical continuous variable states reported. 
Our result is a significant step forward towards continuous variable entanglement purification
which appears to be a straight forward extension of the work presented here \cite{FMFS06}.
Furthermore, in combination with a single non-Gaussian operation (such as single-photon detection) our protocol could be used to purify even states suffering from losses, forming a general CV purification protocol.
Squeezed state purification (or distillation) has also been reported in  \cite{HMDFLLA06}. 
However, our work provides the following distinct features.
First, we considered a rather natural noise source that is indeed expected to be a problem in quantum communication and squeezed mode generation. 
Secondly, our scheme involved \emph{two} copies of a state and it has been proven to work iteratively if more copies of the state are available.

We acknowledge financial support from the Deutsche Forschungsgemeinschaft (DFG), project number SCHN {757/2-1}.
J.F. acknowledges financial support from MSMT (MSM6198959213 and LC06007)
and from the EU under project COVAQIAL (FP6-511004).

\end{document}